\documentclass[epj]{svjour}

\usepackage[dvips]{epsfig}
\usepackage{graphicx}
\newcommand{\be}{\begin{equation}}
\newcommand{\ee}{\end{equation}}
\newcommand{\bea}{\begin{eqnarray}}
\newcommand{\eea}{\end{eqnarray}}

\begin{document}

\title{Multiple stalk formation as a pathway of defect-induced membrane fusion}
\author{D. B. Lukatsky and Daan Frenkel}  
\institute{FOM Institute for Atomic and Molecular Physics,
Kruislaan 407, 1098 SJ Amsterdam, The Netherlands}

\date{Received: date / Revised version: date}

\abstract{
We propose that the first stage of membrane fusion need not be the
formation of a single stalk. Instead, we consider a scenario for
defect-induced membrane fusion that proceeds cooperatively via
multiple stalk formation.  The defects (stalks or pores) attract
each other via  membrane-mediated capillary interactions that
result in a condensation transition of  the defects. The resulting
dense phase of stalks corresponds to the so-called
fusion intermediate.
} 

\maketitle

When two bilayer membranes approach each other, they
may fuse to form a single bilayer membrane. This process (and its
reverse)  are of great importance for many processes  in a living
cell. Nevertheless, the mechanism by which membranes fuse is still
a matter of debate~\cite{Lentz00,Lentz02,Tamm03}.

\begin{figure}[ht]
\begin{center} 
\centerline{\psfig{figure=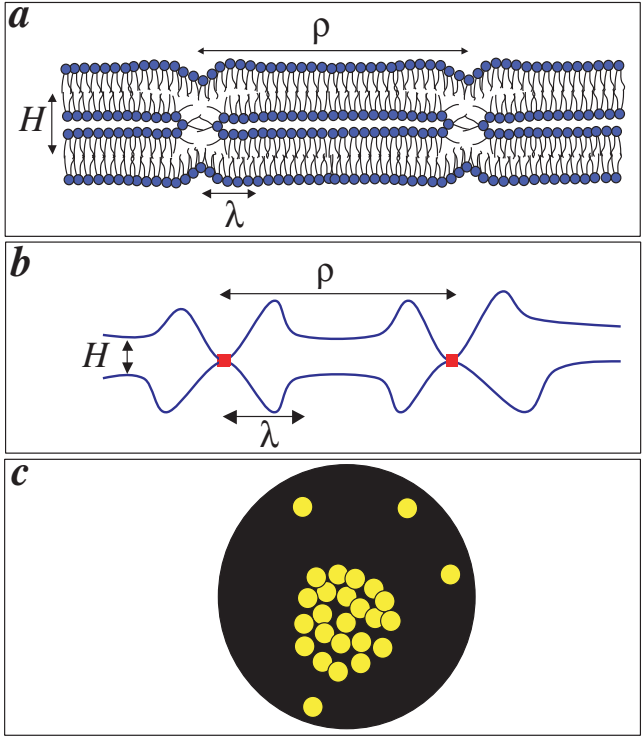,height=10cm,width=8.5cm
}}
\caption{Schematic representation of the cooperative formation of the fusion
intermediate state via condensation of the self-assembled defects (stalks).
The stalks represent elementary defects which interact via an effective, membrane-induced
mechanism. (\textit{a,b}): Two bilayer membranes with two interacting stalks.
(\textit{b}): Within the framework of the model, the membranes have an equilibrium
separation $H$. The stalks constrain undulations of the membranes. The
membranes overshoot the equilibrium separation $H$ in the vicinity of the stalks. This is
the driving factor for the attraction between the stalks. When the stalks are close to each other,
they can share the common overshoot and 
thus reduce the total bending energy. (c): Cartoon of the two-dimensional (2D), phase-separated
(condensed) phase of stalks in the contact zone of two membranes. The dense phase of stalks is equivalent
to a 2D liquid of defects.
The stalks are not the only type of possible defects - pores, or pores coexisting with stalks
can be also formed.}
\label{cartoon}
\end{center}
\end{figure}

The most widely used description of membrane fusion assumes that
it involves several steps. It has been
argued~\cite{Gindell78,Hui81,Kozlov83,Siegel93,Siegel99,Markin02,Kozlov02,Kozlov02a,Kuzmin01}
that the initial connection between the membranes is formed either
through a stalk or through a pore \cite{Siegel93,Kuzmin01}.
In fact, it is now commonly accepted that the
initial inter-membrane contact is, most likely, a stalk
\cite{Lentz02} (see cartoon Fig. \ref{cartoon}.\textit{a}). This initial state is called the hemifusion
\cite{fnTMC}. A recent theoretical analysis of the free-energy
cost of the hemifusion state \cite{Kozlov02} predicts that,
depending on the magnitude of the spontaneous splay of the
lipids, this free energy $F$ can either be positive ($F\approx
45\, k_B\,T$ for the common case of DOPC lipids) or negative
($F\approx -30 \,k_B\,T$ for DOPE lipids that have a large and
negative spontaneous splay).

The second step of the fusion reaction is the expansion of the
hemifusion zone (diaphragm) \cite{Kozlov02a}. The analysis of
ref.~\cite{Kozlov02a} predicts that the expansion of the
hemifusion diaphragm is energetically favorable only if the
spontaneous splay modulus of lipids is large and negative (e.g.,
for DOPE lipids) while it costs energy for membranes composed of
other types of lipids (e.g., DOPC). It is believed that in the
case when the expansion of the hemifusion diaphragm is
energetically unfavorable, fusion proteins generate the additional
driving force needed to expand the diaphragm. The fusion process
is then completed by the subsequent formation and expansion of a
fusion pore.

A direct experimental conformation of the above scenario is still
lacking after more than two decades of investigations. Only
recently \cite{Yang02,Yang03} Yang and Huang have succeeded to
crystallize a stable phase of membrane stalks in
multi-lamellar system, and verify the predicted structure of the
stalk intermediate. No similar experiments have been reported in
the case of two membranes.

Our principal hypothesis in this paper is based on the
observations that (i) two stalks or two pores attract each other,
(ii) the translational entropy associated with the formation of a
single stalk or pore is sufficiently large to allow the
spontaneous emergence of a dilute gas of  such defects, even if
the elastic free energy  of a defect is positive.

We argue below that the defects that are thus formed, will attract
each other and self assemble into a structure that has the
characteristics of a hemifusion zone (see Fig. \ref{cartoon}.\textit{c}). This is a cooperative
effect, similar to conventional order-disorder transition.
We emphasize that the predicted aggregated phase of stalks constitutes an \textit{intermediate}
and not a final stage of membrane fusion. The final stage
can proceed e.g., either through stalk coalescence and a subsequent
pore formation, or through the expulsion of stalks via budding. This final
stage of fusion is beyond the subject of the present work.

We stress that the assumption of multiple stalk formation is not
as farfetched as it may seem. In fact, the transitions from the
lamellar to the inverted hexagonal
phase~\cite{Gruner92,Rappolt03}, and from the  lamellar to sponge
phase~\cite{Sam} are both examples of a similar effect, where the
lamellar phase is transformed into a highly connected structure.
In any event,  the proposed mechanism for the formation of the
intermediate state in the fusion process is not restricted to the
case of spontaneous (passive) multiple stalk formation. Our
conclusions also apply to the case where fusion proteins actively
facilitate stalk formation~\cite{Lentz00,Jena03,Kozlov98}.

To model the condensation of defects during membrane fusion, we
exploit the close analogy between the current process and the
interaction between mobile receptors that are responsible for the
adhesion between cell membranes 
\cite{Bell84,Bruinsma94,Bruinsma00,Bruinsma95,Lipowsky96,Lipowsky01,Komura02,BruinsmaElect,Bruinsma01,deGennes03,Chakraborty01,Kardar03,Chen03,Maier01}.
In what follows, we consider the scenario proposed by Bruinsma,
Goulian, and Pincus in ref. \cite{Bruinsma94}.

\par\noindent 

Consider two bilayer membranes separated by an equilibrium
distance $H$. This equilibrium separation can be a minimum of the
inter-membrane interaction potential $V(h)$
\cite{IsraelachviliBook}, where $h$ is the inter-membrane
separation, or an optimal distance maintained by an external
force. In the case of biological fusion, $H$ is determined by
inter-membrane proteins and by the glycocalyx coating (see ref.
\cite{Lentz00} for a review). We are interested in the
self-assembly of junctions (i.e., stalks or/and pores) between the membranes (see Fig. \ref{cartoon}).
Each junction represents an elementary defect.

We assume that each defect imposes a local membrane spacing $H_0$
that is different from $H$. In the present case of fusion, $H_0$ is simply zero,
as implied by the geometry of a stalk, or a pore~\cite{Kozlov02,Kozlov02a}.
This imposes the boundary
conditions on the inter-membrane spacing at the point of a
junction $\vec{\rho}_0$ \cite{Bruinsma94}: $h(\vec{\rho_0})=0$,
and $\vec{\nabla}h(\vec{\rho}_0)=0$. The free energy of the system
can then be written in the form \cite{Bruinsma94}:
\begin{eqnarray}
F=\int d^2 \vec{\rho} \,\, \left[ \frac{\kappa}{2} (\nabla^2
h)^2+\frac{V^{\prime\prime}(H) }{2} (h-H)^2  \right] \label{fep1}
\end{eqnarray}
where the first term is the Helfrich bending energy
\cite{Helfrich73} with the bending modulus $\kappa$, the second
term is the interaction energy between the membranes;
$h(\vec{\rho}\,)$ depends on the lateral coordinates
$\vec{\rho}=(x,y)$, and $V^{\prime\prime}(H)\equiv
\frac{\partial^2\, V(H)}{\partial h^2}$ [the deviations of the inter-membrane separation $h(\vec{\rho})$
from the equilibrium value $H$ are assumed to be small, and thus
the inter-membrane interaction potential $V(h)$ is expanded to quadratic order in $(h-H)$].

The minimum of Eq. (\ref{fep1}) is given by:
\begin{eqnarray}
\nabla^4h+\frac{h-H}{\lambda^4}=0 \label{mf}
\end{eqnarray}
where $\lambda=[\kappa / V^{\prime\prime}(H)]^{1/4}$ is the
capillary length - the characteristic length of the perturbation
decay. This length is analogous to the stalk width $R$ in the
recent model of the stalk by Kozlovsky and Kozlov \cite{Kozlov02}.
The membrane profile modified by the presence of a junction
located at $\vec{\rho}=0$ is given by the solution of Eq.
(\ref{mf}) with the corresponding boundary conditions, $h(0)=0$,
and $\vec{\nabla}h(0)=0$ \cite{Bruinsma94}:
\begin{eqnarray}
h(\vec{\rho}\,)=H+\frac{4}{\pi}\, H\,
\mbox{kei}(\rho/\lambda) \label{solutionp1},
\end{eqnarray}
where kei($x$) is the Kelvin function. An interesting property of
this profile is that it overshoots the equilibrium inter-membrane
separation, $H$. This effect is even more pronounced if the
non-linear contribution to the interaction potential is taken into
account \cite{BarZiv95,Menes97}. The effect of strong overshooting
of membrane profiles when pinched together by optical tweezers has
been observed experimentally by Bar-Ziv et al. \cite{BarZiv95} and
analyzed theoretically by Menes, Safran, and Kessler
\cite{Menes97a,Menes97}.

The free energy of a single junction is obtained by substituting
the membrane profile $h(\vec{\rho}\,)$ into the free energy Eq.
(\ref{fep1}):
\begin{eqnarray}
F=4\,\kappa\,\frac{H^2}{\lambda^2}.
\end{eqnarray}
This free energy can be directly related to the
free energy of the stalk \cite{Kozlov02}. The dimensionless
parameter $\chi \equiv \frac{H^2}{\lambda^2}$ can thus be obtained
for a given set of the inter-membrane distance $H$ and the stalk
width $\lambda$ ($R$ in the notations of ref. \cite{Kozlov02}).
For example, the free energy of the unconstrained stalk in the
case of DOPC lipids was estimated in ref. \cite{Kozlov02} to be
$F\approx 43\,k_BT$. This gives $\chi \approx 1$, for a typical
value of the bending modulus $\kappa \approx 10\,k_BT$. This
implies that $H/\lambda={\cal O}(1)$. This finding is consistent
with the values for $H$ and $\lambda$ which follow directly from the
calculation of the stalk profile in ref. \cite{Kozlov02}:
$H\approx 6.2\,$nm and $\lambda\approx 8.7\,$nm, leading to the
estimate $\chi\approx 0.7$. In other words, the present, simple
model allows us to interpret the results of the sophisticated
computation of the stalk energy of ref.~\cite{Kozlov02} in terms
of a single parameter $\chi$, that follows directly from the two
intrinsic parameters ($H$ and $\lambda$) of the stalk model. In
fact, it is argued in ref.~\cite{Kozlov02} that  the stalk energy
is minimal when the stalk width $\lambda$ is of the order of the
inter-membrane separation $H$. With that information, it follows
that the free energy of a single stalk $F\approx 4\,\kappa$.

We stress that within our model, a stalk can only have a positive, elastic energy.
This is because we assume that the only effect of a stalk is the constraint on the
inter-membrane separation, and we neglect the topological change upon the stalk formation (see Fig. \ref{cartoon}). 
To take this into account, each monolayer of membranes must be treated separately \cite{fnInclusions}.
This topological change can lead to a negative stalk energy for lipids with sufficiently 
large and negative spontaneous splay \cite{Kozlov02}. Within the framework of our model this 
can be taken into account phenomenologically by considering the stalk free energy $\epsilon_s$ as a sum
of the elastic contribution $F$ and the core free energy $F_{core}$: $\epsilon_s=F+F_{core}$, where
$F_{core}$ can be either positive or negative.

The existence of the attraction between the junctions follows from
the analysis of the free energy of a lattice of junctions \cite{Bruinsma94}. In the limit
$\rho \gg \lambda$, the interaction free energy between two
junctions has the form \cite{fn1}:
\begin{eqnarray}
F_{12}= - C\,\frac{H^2}{\lambda^2}
\,\frac{\kappa}{\sqrt{\rho/\lambda}}\, \exp \left(
-\frac{\rho}{\sqrt{2}\lambda} \right)\,  \sin\left(
\frac{\rho}{\sqrt{2}\lambda} \right), \label{int1}
\end{eqnarray}
where, $C=32\sqrt{2/\pi}\approx 25.5$. The larger the bending rigidity $\kappa$, the stronger the
effective attraction, and the longer the range of this attraction.
The characteristic range of the interaction, $\lambda$ is of the
order of 10-20 nm for an artificial phospholipid membrane under
physiological conditions \cite{Bruinsma00,Bruinsma01} and thus the
long-distance limit Eq. (\ref{int1}) should be accurate if the
junctions are separated by more than 20 nm. We stress that in the
opposite limit $\rho \ll \lambda$, the effective interaction
remains attractive \cite{Bruinsma94}. Hence,  all qualitative conclusions hold
irrespective of the value of $\rho$.

To estimate the phase behavior of the stalks, we use a simple
mean-field model analogous to the ones used to describe the
aggregation of  adhesion molecules or patches within the adhesion
zone of two biological or biomimetic membranes
\cite{Bruinsma00,Bruinsma01,deGennes03,Menes97a}.

The mean-field free energy of the self-assembling junctions can be
constructed with the effective inter-junction potential, Eq.
(\ref{int1}), contributing to the second virial coefficient:
\begin{eqnarray}
f &=& k_B T\,\phi\, \ln \phi + k_B T\,(1-\phi) \ln \,(1-\phi) \\ \nonumber
&+& J\,\phi(1-\phi)+ \epsilon_s \,\phi,
\end{eqnarray}
where $f$ is the free energy per elementary surface cell, $\phi$
is the surface fraction of the junctions, $\epsilon_s$ is the stalk
free energy ($\epsilon_s$ acts as a chemical potential
of defects), and $J$ is the effective, thermodynamic interaction
potential between the junctions. $J$ is obtained on the level of
the linearized second virial coefficient:
\begin{eqnarray}
J=C_1\,\kappa\,\frac{H^2}{\lambda^2}, \label{mf1}
\end{eqnarray}
where $C_1=16\,\sqrt{2}\,\,\pi\,\sin(3\pi/8)\approx 65.7$. Note
that the coefficient $C_1$ is remarkably large. Taking into
account that a typical value of the membrane bending modulus is
$\kappa \sim 10-20\,k_B\,T$, the large value of $C_1$ implies that
the onset of the phase separation of defects occurs at a very low
value of the fusion control parameter, $\chi \equiv
\frac{H^2}{\lambda^2}$. Indeed, it follows from the analysis
of $f$ that the critical point of the ``liquid-gas'' phase
separation of junctions is $J_c/k_BT=2$ and $\phi_c=0.5$. This
implies, if we adopt $\kappa\approx 10\, k_BT$, that the onset of
the phase separation occurs at $\chi_c\approx 0.003$. Comparing
$\chi_c$ with the value of $\chi \approx 1$, estimated above, one
concludes that the phase separation occurs already at a
vanishingly small concentration of defects. The important message
is that the higher the stalk energy, the higher the strength of
the effective, inter-junction attraction, and thus the smaller
concentration of defects induces the phase separation.

The above arguments indicate that stalk condensation should be
possible for reasonable values of the parameters characterizing
biological membranes. Yet, the key question is: does it happen in
practice? In fact, there is experimental evidence that supports
the present scenario:  very recently, Yang and
Huang~\cite{Yang02,Yang03} reported x-ray scattering experiments
that show the spontaneous formation of an ordered, dense
multiple-stalk structure  in a multi-lamellar system of bilayer,
diphytanoyl phosphatidylcholine (DPhPC) lipid membranes. From
their  scattering data, Yang and Huang were able to reconstruct
both the global multiple-stalk structure of the fusing membranes,
and the structure of the individual stalks. The latter shape
turned out to be consistent with the classical stalk structure
\cite{Kozlov02,Kozlov02a,Siegel93,Kuzmin01}. We suggest that this
kind of multiple-stalk structure should be present in practically
all fusion experiments with artificial membranes, provided that
the membranes or vesicles have large enough area of inter-bilayer
contact (see e.g., ref. \cite{MacDonald03}).

There are also recent computer simulations studies
\cite{Schick03,Stevens03}, that report either a simultaneous
formation of two adjacent fusion zones \cite{Stevens03} even in a
fairly small fusing vesicles, or a coexistence of fusion stalks
and pores \cite{Schick03}.

In summary, we propose a possible mechanism of membrane fusion,
via a multiple stalk formation. We predict that the intermediate
stage of membrane fusion represents a phase-separated phase of
self-assembled defects (stalks or pores, or both stalks and
pores). Multiple defect generation is a mechanism alternative to a
hypothesis of a single stalk or pore formation and subsequent
expansion of the fusion diaphragm. The physical origin of the
proposed mechanism is the membrane-induced, effective attraction
between fusion defects, similar to the attraction between adhesion
receptors. Our conclusions apply both in the case of spontaneous
stalk formation and in the case of protein-assisted stalk
formation.

We are grateful to A. Cacciuto and Z. Wang for preliminary simulations of fusion, and to M. Dogterom, B. Mulder, S. Safran,
G. Hed, and M. Kozlov for critical reading of the manuscript. The work of the FOM Institute is part
of the research program of FOM and is made possible by financial
support from the Netherlands organization for Scientific Research (NWO).


\begin{thebibliography}{}
\bibitem{fnTMC} Note that within the framework of the recent model of the talk \cite{Kozlov02},
the hemifusion is equivalent to the trans-monolayer contact (TMC).
\bibitem{fn1} Note, that we obtain the expression for the interaction free energy 
between two junctions which is different by a numerical
factor from the corresponding one reported in ref.
\cite{Bruinsma94}.
\end{thebibliography}
\end{document}